# END-TO-END SPEECH ENHANCEMENT BASED ON DISCRETE COSINE TRANSFORM

*Chuang Geng, Lei Wang*

Beijing University of Posts and Telecommunications, China

## ABSTRACT

Previous speech enhancement methods focus on estimating the short-time spectrum of speech signals due to its short-term stability. However, these methods often only estimate the clean magnitude spectrum and reuse the noisy phase when resynthesize speech signals, which is unlikely a valid short-time Fourier transform (STFT). Recently, DNN based speech enhancement methods mainly joint estimation of the magnitude and phase spectrum. These methods usually give better performance than magnitude spectrum estimation but need much larger computation and memory overhead. In this paper, we propose using the Discrete Cosine Transform (DCT) to reconstruct a valid short-time spectrum. Under the U-net structure, we enhance the real spectrogram and finally achieve perfect performance.

## 1. INTRODUCTION

The goal of speech enhancement is to separate target speech from the background noise to improve the intelligibility and quality of speech. As a fundamental task in signal processing, speech enhancement has a wide range of applications. Such as improving the quality of mobile communications in noisy environments, hearing aids and providing robustness for automatic speech and speaker recognition [1, 2].

Traditional speech enhancement approaches include spectral subtraction [3], Wiener filtering [4], statistical model-based methods [5], and nonnegative matrix factorization [6]. All these models are based on prior knowledge and assumptions of underlying properties of speech and noise, which may not always hold. In recent years, deep learning-based methods have started to attract much attention in the source separation research community by modeling the nonlinear relationship between the mixture and clean speech signals. Typical speech enhancement systems operate in the time Frequency(T-F) domain, only enhancing the magnitude response and leaving the phase in noisy conditions [7]. This may be because there is no clear structure in phase spectrogram, which makes estimating the clean phase from the noisy phase difficult. These methods can be divided into two categories, namely mask based approaches and spectral mapping approaches. Common mask functions include Ideal Binary Mask(IBM) [8] and Ideal Ratio Mask(IRM) [1], which show better performance than direct spectral mapping.

Recently, some research has shown the importance of phase when spectrograms are resynthesized back into time-domain waveforms [9]. One major approach is to use an end-to-end model that takes audio as the raw waveform inputs without using any explicit T-F representation [10, 11]. Since raw waveforms inherently contain phase information, it is expected to achieve phase estimation naturally. Another method is estimating the magnitude spectrum and phase spectrum simultaneously [12-14]. But estimating phase spectrum is not easy, the result in [12] shows that separately enhancing the magnitude response and phase response offers little to no improvement over ratio mask alone. However, later studies that defined a complex IRM(cIRM) to jointly enhance magnitude and phase show better performance than IRM [13]. Recently, [15] proposed elementary complex building blocks for complex-valued deep neural networks, which put all arithmetic in the complex domain. Based on these blocks, [14] proposed a new architecture which combined the advantages of both deep complex networks and the U-net [18], and finally achieving state-of-the-art performance.

## 2. RELATION TO PRIOR WORK

Recently deep learning-based method seems more suitable for speech enhancement due to its nonlinear model capability. Many studies have attempted to solve the phase estimation problem because of its importance and difficulty. The research in [16] tries to minimize a loss defined in the frequency domain to solve the problem that the combination of noisy phase and estimated magnitude is unlikely a valid STFT. But the result shows that this frequency loss using both the magnitude and phase information does not give an as good performance as using only the magnitude information, which is unexpected. Another method tries to define a cIRM that jointly enhances the magnitude and phase spectrum of noisy speech [13]. However, it only lets the mask in the complex domain, while the whole network structure is real-valued, which could not represent spectral patterns. Based on complex component proposed by [15], [14] and [17] separately combine it with U-net [18] and

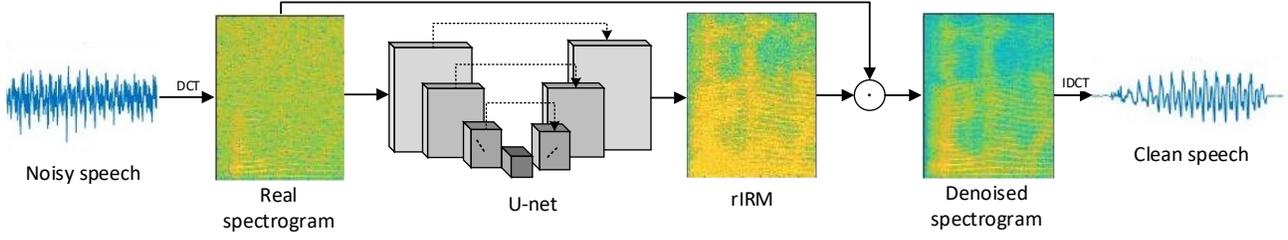

**Fig.1**.Illustration of speech enhancement framework with the U-net

Feed-forward network. These methods are reasonable therefore achieving the best performance. However, complex blocks such as complex convolution and complex batch normalization all need much more computation and memory overhead than that of real components. Besides, complex batch normalization reduces the correlation between the real part and the imaginary part of the complex spectrum. In theory, we could recover the complex spectral only from its real part or imaginary part after symmetric transformation, which could prove the internal relations between them.

To escape the phase estimation problem, we use DCT to reconstruct a real-valued short-term spectrum. Then we use the common real-valued U-net to estimate the clean spectrum. Under this structure, we finally achieve comparable performance to complex networks.

## 3. SYSTEM OVERVIEWS

Given the input sequence, we first use DCT to extract the real spectrogram. Next, we treat each real spectrogram as the input to the U-net architecture. At the output of the U-net, we estimate the mask function and then multiply it to the noisy input to achieve the enhanced spectrogram. Finally, we use inverse DCT to recover the estimated clean speech waveform. The whole framework is shown in Fig.1. Below we will detail our approach, starting with the theory of DCT, followed by the design of the U-net structure. Finally, we will introduce the mask function.

### 3.1.Discrete cosine transform

We have proposed to derive Conjugate Symmetric Sequence $x_e(n)$ from the original signal $x(n)$ to solve the phase estimation problem, where $x_e(n)$ is defined as

$$x_e(n) = \frac{1}{2}[x(n) + x^*(-n)] \quad (1)$$

It can be proved that the discrete-time Fourier transform (DTFT) of $x_e(n)$ is real-valued and equals to the real part of the complex spectrum of $x(n)$. Then we put the real spectrogram into the network to estimate the enhanced spectrogram. Finally, we could recover $x(n)$ from $x_e(n)$ in the time domain. In later studies, we found that the DTFT of $x_e(n)$ is theoretically equivalent to the DCT of $x(n)$, which we could obtain the real spectrum in an easier way.

The DCT of $x(n)$ is defined as

$$X_c(k) = \sqrt{\frac{2}{N}}\beta(k)\sum_{n=0}^{N-1}x(n)\cos(\frac{\pi k(2n+1)}{2N}) \quad k=0,1,\cdots,N-1 \quad (2)$$

and its inverse DCT is

$$x(n) = \sqrt{\frac{2}{N}}\sum_{k=0}^{N-1}\beta(k)X_c(k)\cos(\frac{\pi k(2n+1)}{2N}) \quad n=0,1,\cdots,N-1 \quad (3)$$

where

$$\beta(k) = \begin{cases} \frac{1}{\sqrt{2}} & k=0 \\ 1 & 1 \leq k \leq N-1 \end{cases} \quad (4)$$

Because DCT and inverse DCT only operate in the real domain, $X_c(k)$ is also real-valued.

In the definition of DCT, there contain N cosine sequences which are

$$c_k = \sqrt{\frac{2}{N}}\beta(k)\cos(\frac{\pi k(2n+1)}{2N}) \quad k=0,1,\cdots,N-1 \quad (5)$$

It can be proved that these N sequences are orthogonal to each other, constructing a standard orthogonal basis. So DCT is an orthogonal transformation defined on this orthogonal basis. We treat each cosine sequence as a row vector and define the DCT matrix as

$$W_{DCT} = \begin{bmatrix} c_0 \\ c_1 \\ \vdots \\ c_{N-1} \end{bmatrix} \quad (6)$$

From the definition of the $W_{DCT}$, we could receive that

$$W_{DCT} \times W_{DCT}^T = W_{DCT}^T \times W_{DCT} = I \quad (7)$$

So the matrix represents of DCT can be written as

$$X_c = W_{DCT}x \quad (8)$$

$$x = W_{DCT}^{-1}X_c = W_{DCT}^T X_c \quad (9)$$

A close relationship between DCT and DFT can be found by defining a 2N points real sequence $x_{es}(n)$:

$$x_{es}(n) = x((n))_{2N} + x((-n-1))_{2N} \quad (10)$$

Where $x((n))_{2N}$ represents a cyclic shift operation. The relationship between $x(n)$ and $x_{es}(n)$ is in Fig.2. It shows that $x_{es}(n)$ is the even symmetric sequence of $x(n)$. We

denote $X_{es}(k)$ as the 2N points DFT of $x_{es}(n)$, when $0 \leq k \leq N-1$, we have

$$X_{es}(k) = \frac{\sqrt{2N}}{\beta(k)} e^{-j\frac{\pi k}{2N}} X_c(k) \quad (11)$$

This shows that $X_c(k)$ is the spectrum of $x_{es}(n)$, which is the even symmetric sequence derived from $x(n)$.

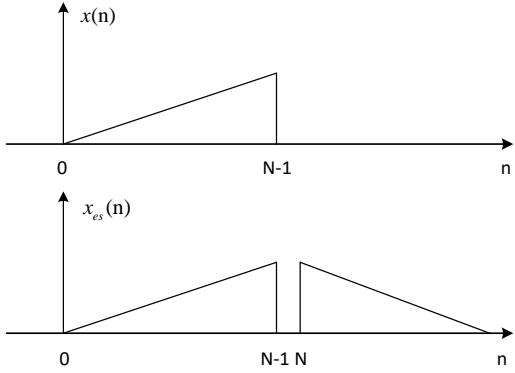

**Fig.2**. The relationship between $x(n)$ and $x_{es}(n)$

### 3.2. U-net structure

The U-net is a well-known architecture composed as a convolutional autoencoder with skip-connections, originally proposed for medical imaging in the computer vision community [18]. The U-net consists of two stages, namely the encoder stage and the decoder stage. In the encoder stage, we use strided convolution to realize the subsample and in decoder using strided transpose convolution to realize the upsample. After each convolutional layer, we use batch normalization to normalize the layer output. For the activation function, we choose parametric RELU [19] whose negative x-axis slope is trainable:

$$f(x) = \begin{cases} x & x > 0 \\ ax & x \leq 0 \end{cases} \quad (12)$$

Description of encoder and decoder block is in Fig.3:

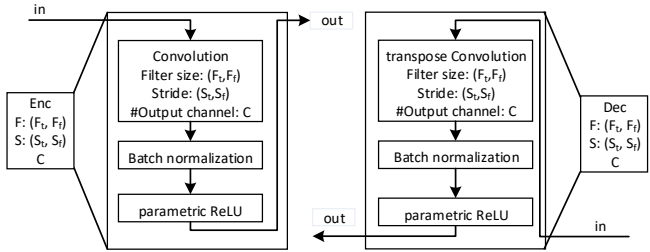

**Fig.3**. Description of encoder and decoder blocks. $F_t$ and $F_f$ denote the convolution filter size along the time and frequency axis, respectively. $S_t$ and $S_f$ denote the stride size of the convolution filter. C denotes the number of output channels.

### 3.3. Real-valued mask function

Although it is possible to directly estimate the spectrogram of a clean speech signal, it has been shown that better performance can be achieved by applying a weighting mask to the mixture spectrogram [10]. The definition of mask function is

$$M_{t,f} = \frac{S_{t,f}}{Y_{t,f}} \quad (13)$$

Where $S_{t,f}$ and $Y_{t,f}$ denote the clean speech spectrogram and noisy spectrogram in a particular T-F unit respectively. Because $S_{t,f}$ and $Y_{t,f} \in \mathbb{R}$, meaning that $M_{t,f}$ is in the range $(-\infty, +\infty)$. we call this type of mask function rIRM. It is difficult for the network to optimize from an infinite search space compared to a bounded one. In magnitude spectrogram estimation, we usually choose a sigmoid function to restrict the IRM to the range [0, 1]. Here, we design a scaled tanh activation function to estimate the rIRM:

$$rIRM = K \frac{1 - e^{-C \cdot M_{t,f}}}{1 + e^{-C \cdot M_{t,f}}} \quad (14)$$

Where rIRM is restricted in the range [-K, K], and C is used to control the steepness.

## 4. EXPERIMENTS SETUP

### 4.1. Dataset

Voice Bank-DEMAND[20]: This dataset has been used in recent several denoising works which we choose as baselines. To generate the training set, we choose 28 speakers (14 male and 14 female) from the Voice Bank corpus [21] and 10 types of noise data from the DEMAND [22]. The signal-to-noise (SNR) values used for training were: 15dB, 10dB, 5dB and 0dB. The test set was chosen from two other speakers (one male and one female), mixed with 5 other noise types. The SNR values for test set are 17.5dB, 12.5dB, 7.5dB and 2.5dB. We resample them to 16kHz and normalize the wave amplitude to [-0.5, 0.5].

### 4.2. Experimental setups

Given the input sequence, we first frame each signal using the hamming window of size 1024 with a frame shift of 64 samples. Next, using DCT to extract the real spectrogram, which is the input of the U-net. The structure of the U-net is in Fig.4. It has ten blocks: five encoder layers and five decoder layers. We add skip-connection to each layer. The convolution kernels are set to be independent of each other by initializing the weight tensor as an orthogonal matrix. After each convolution layer, batch normalization and parametric ReLU are used. In the very last layer the batch normalization is not used and mask function is applied instead. We choose K=2 and C=0.5 for rIRM. We train the network by Adam [23] optimizer where the learning rate is 1e-3 and beta1 and beta2 are 0 and 0.999, respectively. Epsilon is set to 1e-8 for numerical stability. The batch size is 16. The loss function is the same as in [14].

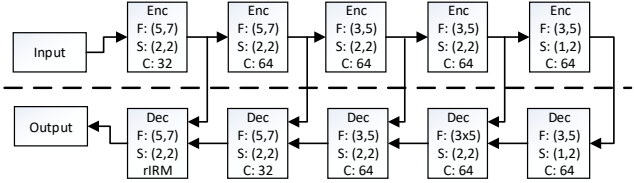

**Fig.4**.Discription of the U-net structure, the encoder and decoder blocks are described in Fig.3.

### 4.3.Experimental results

For a fair model capability comparison, we choose SEGAN [11], Wavenet [12] and DCUnet [14] as the baselines because they use the same Voice Bank-DEMAND corpus. The SEGAN is based on Generative Adversarial Network whose Generator is the U-net structure. The Wavenet denoising network is modified on [24], which is a Generative model for raw audio synthesis. The SEGAN and Wavenet are end-to-end structures directly denoising on the raw waveform. The DCUnet is the state-of-the-art denoising network that combines both the U-net and deep complex networks. For a fair comparison, we choose DCUnet-10(RMRn) and DCUnet-10(cRMCN). The network of RMRn is absolutely the same as ours, but its input is magnitude spectrogram and the ground truth phase was given during training and test. While the cRMCN is a complex-valued network, whose channels per layer are $\sqrt{2}$ times that of RMRn and ours.

The evaluation metrics we used are Perceptual evaluation of speech quality (PESQ) [25] and three other composite scores, which are CSIG for signal distortion, CBAK for background noise intrusiveness and COVL for overall signal quality [26]. PESQ values in the range [-0.5, 4.5] and the other three in [1, 5]. All metrics with higher scores mean better performance. The comparison results show in Table 1. We see that our method is better than SEGAN and Wavenet by a large margin. Compared with RMRn, our method performs better in all metrics and could achieve comparable performance to cRMCN. However, compared with the complex-valued network, our real-valued network is simpler and requires less computation and memory overhead.

**Table 1**.Evaluation results with different methods

|  | PESQ | CSIG | CBAK | COVL |
|---|---|---|---|---|
| Noisy | 1.97 | 3.35 | 2.44 | 2.63 |
| SEGAN | 2.16 | 3.48 | 2.94 | 2.8 |
| Wavenet |  | 3.62 | 3.23 | 2.98 |
| DCUnet-10(RMRn) | 2.51 | 3.71 | 3.23 | 3.01 |
| DCUnet-10(cRMCN) | **2.72** | 3.74 | **3.6** | 3.22 |
| Ours | 2.7 | **3.9** | 3.29 | **3.29** |

### 4.4.Multiple noise test

In this section, we test a situation where a speech signal is contaminated with multiple noises. For comparison, we use the Wiener filter method as our baseline. First, we add blue, pink, violet, white noise on a clean utterance selected from the TIMIT database sequentially. The SNR value is 10dB. Then we use both the Wiener filter and our model to filter the noisy speech. The filter results are shown in Fig.5.

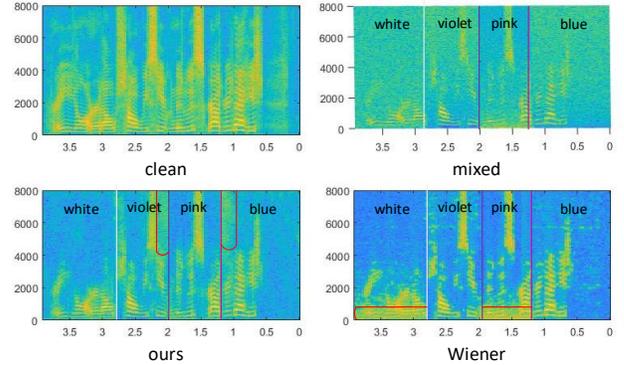

**Fig.5**.Multiple noises test results. (a) is the clean speech spectrogram. (b) is the noisy speech spectrogram, blue, pink, violet and white noises are added sequentially. (c) is the denoised spectrogram based on our model. (d) is the denoised spectrogram based on the Wiener filter.

As can be seen from the picture (d), the Wiener filter perfectly filters out blue and violet noise, as well as white noise at high frequencies, but can not filter out pink noise and low-frequency white noise. This is because the Wiener filter estimates the noise spectrum from the blue noise, whose spectrum is mainly at high frequencies. From the picture (c), we could see that our model could denoise all types of noise. However, at the switching point of blue/violet noise and pink noise (the area in the red box), our model can not filter the noise effectively. This may be because the noisy spectrum is not continuous in that place. Next, we will try to solve the problem of noise tracking in speech enhancement.

## 5. DISCUSSION AND CONCLUSION

In this paper, we proposed using DCT to reconstruct a valid short-term spectrogram in the real domain, successfully escaping the phase estimation problem in speech enhancement. The experiment result shows that our method could achieve comparable performance to DCUnet. Besides, we test a situation where a speech signal is contaminated with multiple noises. And the result shows that at switching point of noise types, our method could not filter the noise effectively. Next, we will address the noise tracking problem in speech enhancement.

https://github.com/BYRTIMO/END-TO-END-SPEECH-ENHANCEMENT-BASED-ON-DISCRETE-COSINE-TRANSFORM